\documentstyle[prabib,aps,preprint,psfig]{revtex}
\date{today}
\begin{document}

\noindent
%%%%%%%%%%%%%%%%%%%%%%%%%%%%%%%%
\begin{center}{\Large \bf 
Molecular-orbital structure in neutron-rich C isotopes} \\

\vskip.25in
{\it N. Itagaki,$^1$\footnote{E-mail: itagaki@phys.s.u-tokyo.ac.jp}
S. Okabe,$^2$ K. Ikeda,$^3$ and I. Tanihata$^3$}

{$^1$ \it 
Department of physics, University of Tokyo, Hongo, Tokyo 113-0033, Japan
}

{$^2$ \it Center for Information and Multimedia Studies,
Hokkaido University, Sapporo 060-0810, Japan}

{$^3$ \it 
The Institute of Physical and Chemical Research (RIKEN),
Wako, Saitama 351-0198, Japan}

\end{center}

\begin{abstract}                % DON'T CHANGE THIS LINE
The molecule-like structure of the 
C isotopes (A=12, 14, 16) is investigated 
using a microscopic $\alpha$+$\alpha$+$\alpha$+$n$+$n$+$\cdot \cdot \cdot$
model. The valence neutrons are classified 
based on the molecular-orbit (MO) model, and both $\pi$-orbit and 
$\sigma$-orbit are introduced around three $\alpha$-clusters.
The valence neutrons which occupy the $\pi$-orbit increase
the binding energy and stabilize the linear-chain of 3$\alpha$
against the breathing-like break-up. 
However, $^{14}$C with the $\pi$-orbit does not show clear energy
minimum against the bending-like path.
The combination of 
the valence neutrons in the $\pi$- and the $\sigma$-orbit 
is promising to stabilize the linear-chain state against the
breathing- and bending- modes,
and it is found that 
the excited states of $^{16}$C with 
the $(3/2^-_\pi)^2(1/2^-_\sigma)^2$ configuration for the four
valence neutrons is 
one of the most promising candidates for such structure.
\end{abstract}
\begin{center}
PACS number(s): 21.10.-k, 21.60.Gx
\end{center}

\section{INTRODUCTION}
A survey of the molecule-like structure is one of the
most challenging subjects in light neutron-rich nuclei.
In Be isotopes, recently, decay into fragments of He isotopes
($^4$He, $^6$He, $^8$He) has been observed 
from the excited states of $^{10}$Be\cite{Soi} and $^{12}$Be\cite{Kor,Freer},
and the presence of a two-center configuration is suggested.
From the theoretical side, these states are studied by
various models by which 
the molecular-orbital nature of the weakly bound neutrons 
around the two $\alpha$-clusters
has been revealed \cite{Okabe-S,Seya,EnyoBe10,Ogawa,Ita}.

As for multi-cluster configurations in light $\alpha$-nuclei
beyond the two-center systems, 
the existence of $N\alpha$ states has been predicted around the threshold energy
in the so-called Ikeda diagram \cite{Ikeda68}. 
For example,
it has been suggested that the second $0^+$ state of $^{12}$C
has a 3$\alpha$-like molecular configuration\cite{Morinaga}.
However, according to many theoretical analyses\cite{Suppl68}, 
the state 
is not necessary to have a linear-chain of
3$\alpha$, but is described as a weak-coupling state with a triangular shape
or the $^8$Be+$\alpha$ configuration.
Furthermore, a rotational band of $^{16}$O
with a very large moment of inertia has been observed
around the 4$\alpha$ threshold energy region\cite{Cheva},
and the 4$\alpha$ linear-chain have been 
discussed \cite{Horiuchi72,Rae}.
However, the 4$\alpha$-chain has not been experimentally confirmed yet.

Recently, the discussions of the well-developed cluster structure 
are extended to the neutron-rich nuclei, and the role of valence 
neutrons which stabilize the linear-chain structure has been pointed out.
For example, von Oertzen has 
extended his analyses for the molecular structure in Be isotopes
\cite{Oertzen} to C isotopes,
and the linear-chain state consisting of 3$\alpha$ and
valence neurons around it has been speculated.
Even if the 3$\alpha$-system without valence neutrons ($^{12}$C) 
does not have a linear-chain structure,
the valence neutrons around it
are expected to increase the binding energy and stabilize
the linear-chain state.

In this paper, the molecular-orbital
approach introduced for the Be isotopes is applied to 
a study of C isotopes (A=12, 14, 16).
The appearance of prolonged structure with the $\alpha$-$\alpha$
core in $^{10}$Be\cite{Ita} and $^{12}$Be\cite{Ita2}
has been successfully described by 
the $\alpha$+$\alpha$+n+n+$\cdot \cdot$ model, where the orbits for the
valence neutrons are classified based on 
the molecular-orbit (MO) model\cite{Abe}.
In the present study, MO is extended to C isotopes,
we focus on the stabilization of the linear-chain 
structure with the 3$\alpha$-core
to which valence neutrons are added.

\section{Model} 
The total wave function of a microscopic 
$\alpha$+$\alpha$+$\alpha$+$n$+$n \cdot \cdot \cdot$ model 
is fully antisymmetrized and
expressed by a superposition of basis states
centered to different relative distances between 
the $\alpha$-clusters ($d$)
with various configurations of the valence neutrons 
($c1$, $c2$ $\cdot \cdot$) around the $\alpha$-clusters:
\begin{equation}
\Phi^J_{MK}=\sum_{d,c1,c2 \cdot \cdot}
P^J_{MK}
{\cal A}[\phi^{(\alpha)}_1\phi^{(\alpha)}_2 \phi^{(\alpha)}_3
(\phi^{c1}_1\chi_1)(\phi^{c2}_2\chi_2)\cdot \cdot \cdot].
\end{equation}
The projection to the eigen-states of angular momentum $(P^J_{MK})$ 
is performed numerically.
Each $\alpha$ cluster consisting of four nucleons
is described by Gaussians ($G_{\alpha i}$) centered at $R_{\alpha i}$ 
and spin-isospin wave function$(\chi)$: 
\begin{equation}
\phi^{(\alpha)}_i=
G_{R_{\alpha i}}^{p\uparrow}
G_{R_{\alpha i}}^{p\downarrow}
G_{R_{\alpha i}}^{n\uparrow}
G_{R_{\alpha i}}^{n\downarrow}
\chi_{p\uparrow}
\chi_{p\downarrow}
\chi_{n\uparrow}
\chi_{n\downarrow}, \ \ \ \ \ \ i=1, 2, 3.
\end{equation}
\begin{equation}
G_{R} = \left( {2\nu \over \pi}\right)^{3 \over 4}
\exp[-\nu(\vec r-\vec R)^2],
\ \ \ \ \ \ \nu=1/2\beta^2,
\end{equation}
where, the oscillator parameter $(\beta)$ is equal to 1.46 fm.
For the linear-chain state, the values of
$\{ R_{\alpha i} \}$ are $-d$, $0$, and $+d$ on the $z$-axis.
Each valence neutron ($\phi^{ci}_i\chi_i$) around 
the $\alpha$-$\alpha$-$\alpha$ core
is expressed by a linear combination of local Gaussians:
\begin{equation}
\phi^{ci}_i\chi_i =
\sum_{j} g_{j} G_{R_j} \chi_i.
\end{equation}

These valence-neutron orbits are classified according to
the MO picture\cite{Abe}.
The orbit of the valence neutron perpendicular 
to the $z$-axis of the 3$\alpha$ linear-chain core
is called $\pi$-orbit, 
and one along the $z$-axis is called $\sigma$-orbit.
The antisymmetrization imposes the forbidden space
for the valence neutrons;
the $\pi$-orbit must have at least
one node perpendicular to the $z$-axis,
and the $\sigma$-orbit must have at least three
nodes since three $\alpha$-clusters along the $z$-axis
already occupy the orbitals with $n_z = 0, 1, 2$. 
The amplitude 
of the $\pi$- and the $\sigma$-orbit
are schematically drawn in Fig. 1. 
%\noindent
%\begin{center} 
%--------------\\
%Fig. 1 \\
%--------------\\
%\end{center}

In the present framework,
each valence-neutron orbit is 
introduced to have a definite $K^\pi$ value
at the zero limit of centers of local Gaussians 
$(\{ R^j \})$ describing the spatial distribution of the orbit. 
The precise positions of $\{ R^j \}$
are determined variationally 
before the angular-momentum projection.
Since the values of $\{ R^j \}$ are optimized to be finite,
the orbits are not exactly the eigen-state of $K^\pi$,
and are labeled as $\bar K^\pi$.
For the $\pi$-orbit with $\bar K^\pi=3/2^-$ ($| 3/2^-_\pi \rangle$),
the spatial part and the spin part of $\bar K$ 
are defined to be parallel ($rY_{11}|n\uparrow \rangle$), 
for which the spin-orbit interaction acts attractively.
At the same time, $| 3/2^-_\pi \rangle$ is described
as a linear combination of two orbits 
centered at right- and left-hand side of the system
based on the MO picture.
\begin{equation}
| 3/2^-_\pi \rangle
={1 \over \sqrt{N_\pi}}
\{
{1 \over \sqrt{2}}(p_x+ip_y)_{+a}+
{1 \over \sqrt{2}}(p_x+ip_y)_{-a}
\}|n\uparrow \rangle.
\end{equation}
\begin{equation}
(p_x)_{\pm a}=G_{\pm a \vec e_z+b\vec e_x}-G_{\pm a \vec e_z-b\vec e_x}, \ \ \
(p_y)_{\pm a}=G_{\pm a \vec e_z+b\vec e_y}-G_{\pm a \vec e_z-b\vec e_y}.
\end{equation}
Here, $(p_x+ip_y)_{\pm a}$ denotes
the $p$-orbit centered at $\pm a$ on the $z$-axis,
and these variational parameters $a$ and $b$ are optimized by using 
the Cooling Method in antisymmetrized molecular dynamics 
(AMD) \cite{Enyo95b,Ono92} for each basis state.
Furthermore, the $| 1/2^-_\pi \rangle$ orbit 
where the spin-orbit interaction acts repulsively
can also be defined by changing the spin-direction of $| 3/2^-_\pi \rangle$,
where the spatial part and the spin part of $\bar K$ 
are anti-parallel ($rY_{11}|n\downarrow \rangle$): 
\begin{equation}
| 1/2^-_\pi \rangle
={1 \over \sqrt{N_\pi}}
\{
{1 \over \sqrt{2}}(p_x+ip_y)_{+a}+
{1 \over \sqrt{2}}(p_x+ip_y)_{-a}
\}|n\downarrow \rangle.
\end{equation}
The distribution of the
$\sigma$-orbit is just along the 3$\alpha$ axis,
then it is introduced to have three nodes.
$| 1/2^-_\sigma \rangle$ is represented as a linear combination of three 
orbits with $\bar K^\pi=1/2^-$,
whose centers are $+a$, $0$, and $-a$ on the $z$-axis.
\begin{equation}
| 1/2^-_\sigma \rangle
={1 \over \sqrt{N_\sigma}}
\{(p_z)_{+a}-(p_z)_{0}+(p_z)_{-a}\}|n\uparrow \rangle,
\end{equation}
\begin{equation}
(p_z)_{\pm a}=G_{a\vec e_z+b\vec e_z}-G_{a\vec e_z-b\vec e_z}, \ \ \
(p_z)_0=G_{b\vec e_z}-G_{-b\vec e_z}.
\end{equation}
These three orbits
($ | 3/2^-_\pi \rangle$, $| 1/2^-_\pi \rangle$ and $| 1/2^-_\sigma \rangle$)
are the basic building blocks for the molecular-orbital
structure. 
Also,
$| -3/2^-_\pi \rangle$, 
$| -1/2^-_\pi \rangle$ 
and 
$| -1/2^-_\sigma \rangle$ 
orbits
are introduced by taking the time reversal
of $| 3/2^-_\pi \rangle$, $| 1/2^-_\pi \rangle$,
and $| 1/2^-_\sigma \rangle$ orbits, respectively. 

The Hamiltonian 
and the effective nucleon-nucleon interaction
are the same as in Refs.\cite{Ita,Ita2}, 
and parameters of Volkov No.2\cite{VolkovInt} for the central part
and the G3RS spin-orbit term\cite{G3RS} 
for the spin-orbit part are 
determined 
from the $\alpha+n$ and $\alpha+\alpha$ 
scattering phase shifts, and the binding energy of
deuteron is also reproduced with these parameters.

\section{Results and discussions for bending stability} 
In the following part, we show the calculated results for the 
stability of the linear-chain state for various configurations.
The isotopes and configurations which we take into account
are $^{12}$C, $^{14}$C$(3/2^-_\pi)^2$
(two $n$'s in the $\pi$-orbits),
$^{14}$C$(1/2^-_\sigma)^2$
(two $n$'s in the $\sigma$-orbits),
$^{16}$C(($3/2^-_\pi)^2 (1/2^-_\pi)^2)$
(four $n$'s in the $\pi$-orbits) and
$^{16}$C(($3/2^-_\pi)^2 (1/2^-_\sigma)^2)$
(two $n$'s in the $\pi$-orbits and two $n$'s in the $\sigma$-orbits). 
As schematically shown in Fig. 2, two variational paths
are introduced corresponding to the
breathing-like (Fig. 2 (a)) 
and the bending-like (Fig. 2 (b)) degrees of freedom.
The parameters $d$ and $\theta$ stand for 
the $\alpha$-$\alpha$ distance
and the bending angle of the 3$\alpha$-core, respectively.

Firstly, we show the 
$0^+$ energy curves for the linear-chain structure against
the breathing-path in Fig. 3.
It is found that the
energy pocket around $d = 3$ fm becomes deeper
as the increase of number of valence neutrons in the $\pi$-orbit
($^{12}$C $\to$ $^{14}$C$(3/2^-_\pi)^2$ $\to$
$^{16}$C(($3/2^-_\pi)^2 (1/2^-_\pi)^2)$).
%
%\noindent
%\begin{center}
%--------------\\
%Fig. 3 \\
%--------------\\
%\end{center}
Using our framework, the binding energy of one $\alpha$-cluster
is calculated to be 27.5 MeV, then the $\alpha$+$\alpha$+$\alpha$
+$n$-neutrons threshold energy is $-82.5$ MeV.
The 3$\alpha$-system ($^{12}$C) 
has minimal energy around $d = 3.5$ fm,
however, this is too shallow
to conclude the stability of the linear-chain state.
On the contrary, 
in $^{14}$C$(3/2^-_\pi)^2$,
there appears evident minimal energy around $d = 3$ fm.
The energy 
($\sim-82$ MeV)
is lower than $^{12}$C by 11 MeV
and the energy pocket is much deeper.
This energy corresponds to the excitation energy of 18 MeV 
from the ground state calculated 
with an equilateral-triangle configuration for the 3$\alpha$-core,
which is $-101.2$ MeV.
The $^{16}$C$((3/2^-_\pi)^2(1/2^-_\pi)^2)$ 
configurational state
is most stable 
among states studied and
has an energy pocket of $\sim-86$ MeV, where
the $\alpha$-$\alpha$ distance is $d = 2.5$ fm, shorter than 
those for $^{12}$C and $^{14}$C$(3/2^-_\pi)^2$.
Therefore, the $\pi$-orbit is found to stabilize the linear-chain
structure as the increase of valence neutrons
($^{12}$C $\to$ $^{14}$C$(3/2^-_\pi)^2$ $\to$
$^{16}$C(($3/2^-_\pi)^2 (1/2^-_\pi)^2)$).

Next, we discuss the case
where the valence neutrons occupy the $\sigma$-orbit. 
$^{14}$C$(1/2^-_\sigma)^2$
has an excitation energy higher by about 14 MeV 
in comparison with $^{14}$C$(3/2^-_\pi)^2$.
It is rather surprising that the difference 
is only 14 MeV in spite of the fact that $3/2^-_\pi$ has only one node
and $1/2^-_\sigma$ has three nodes.
This is because the $\sigma$-orbit
is along the $\alpha$-$\alpha$-$\alpha$ core:
The higher nodal orbits along the symmetry axis 
become low-lying as a result of the clustering of the core.
The $\sigma$-orbit enhances the prolonged shape of the 3$\alpha$-core,
and the optimal $d$-value is $\sim$3.5 fm.
However, the $^{14}$C$(1/2^-_\sigma)^2$
case has no deep pocket enough to be stabilized
against the breathing-like path.
When two more valence neutrons occupy the $\pi$-orbit,
although this minimal energy is higher by 5 MeV than that of
$^{16}$C$((3/2^-_\pi)^2(1/2^-_\pi)^2)$, 
$^{16}$C$((3/2^-_\pi)^2(1/2^-_\sigma)^2)$
has the minimal $0^+$ energy of $\sim-77$ MeV.
The calculated energy pocket is deep enough to
guarantee the stability for the breathing-like path.
Therefore, it is summarized that
the linear-chain structure cannot be
stabilized against the breathing-like mode 
with the $\sigma$-orbit only 
in $^{14}$C$((1/2^-_\sigma)^2)$, however,
the neutrons in $\pi$-orbits
increase the binding energy and
prevent the break-up of the system
in $^{14}$C$((3/2^-_\pi)^2)$, 
$^{16}$C$((3/2^-_\pi)^2(1/2^-_\pi)^2)$, and 
$^{16}$C$((3/2^-_\pi)^2(1/2^-_\sigma)^2)$.

Finally, the stability of these linear-chain states
against the bending-like path is examined.
The $0^+$ energy curves of $^{12}$C,
$^{14}$C$((3/2^-_\pi)^2)$,
$^{14}$C$((1/2^-_\sigma)^2)$,
$^{16}$C$((3/2^-_\pi)^2(1/2^-_\pi)^2)$,
and
$^{16}$C$((3/2^-_\pi)^2(1/2^-_\sigma)^2)$
against the $\theta$-value
are shown in Fig. 4. 
%
%\noindent
%\begin{center}
%--------------\\
%Fig. 4 \\
%--------------\\
%\end{center}
Except for the case of 
$^{16}$C$((3/2^-_\pi)^2(1/2^-_\sigma)^2)$,
the curvature of these states is rather monotonic
and the energy minimum does not clearly appear.
In $^{14}$C, 
the orthogonality 
between the linear-chain configuration and low-lying states
with equilateral-triangle configuration 
of 3$\alpha$ is taken into account.
This effect works but 
not sufficiently to push up the energy
of the state with a finite bending angle.
However, as clearly seen in Fig. 4, 
the $^{16}$C$((3/2^-_\pi)^2(1/2^-_\sigma)^2)$
case shows a sharp increase of the $0^+$ energy
as the increase of the bending angle
and 
is found to be stable against the bending-like path.
This feature is much different from 
$^{12}$C, $^{14}$C$((3/2^-_\pi)^2)$, 
$^{14}$C$((1/2^-_\sigma)^2)$,
and $^{16}$C$((3/2^-_\pi)^2(1/2^-_\pi)^2)$ cases. 
From the analysis above,
the linear-chain configuration can be stabilized 
against the breathing-like path by neutrons
in the $\pi$-orbit ($^{14}$C$((3/2^-_\pi)^2)$,
$^{16}$C$((3/2^-_\pi)^2(1/2^-_\pi)^2)$, 
$^{16}$C$((3/2^-_\pi)^2(1/2^-_\sigma)^2)$), but
$^{16}$C$((3/2^-_\pi)^2(1/2^-_\pi)^2)$
is only the case which is stable also against 
the bending-like path.

We further discuss 
the reason for the sharp increase of the $0^+$ energy in
$^{16}$C$((3/2^-_\pi)^2(1/2^-_\sigma)^2)$ 
against the bending-like path.
The $0^+$ energy increases by 15.7 MeV 
from $\theta = 0^o$ to $\theta = 30^o$,
in which the kinetic energy
part is 10.3 MeV. 
To understand the energy increase with the increase of bending
angle $\theta$ of this case,
we calculate and compare the overlap 
between the wave functions with $\theta = 0^o$ and $\theta = 30^o$
for various configurations in Table I, where the results are shown.
In $^{12}$C, 
the wave functions with $\theta = 0^o$ and $\theta = 30^o$
have the squared overlap of 0.91, and 
$^{14}$C$((3/2^-_\pi)^2)$ has almost the same value.
$^{14}$C$((3/2^-_\sigma)^2)$ has the value of 0.85, smaller
than $^{14}$C$((3/2^-_\pi)^2)$ by only 6 $\%$, and 
$^{16}$C$((3/2^-_\pi)^2(1/2^-_\pi)^2)$ has almost the same value
as the $^{14}$C$((1/2^-_\sigma)^2)$ case.
This result shows that the overlaps additionally decreases
a little for the $\sigma$-orbital neutrons, 
and also for the $\pi$-orbital
neutrons as the increase of the valence neutrons.
In spite of these, the overlap between 
$\theta = 0^o$ and $\theta = 30^o$
for $^{16}$C$((3/2^-_\pi)^2(1/2^-_\sigma)^2)$
case shows a significantly large decrease to 0.60. 
$^{16}$C$((3/2^-_\pi)^2(1/2^-_\sigma)^2)$
is only the configuration which shows drastic decrease
of the overlap between $\theta = 0^o$ and $\theta = 30^o$.

As discussed in following part,
it can be known that the drastic decrease 
as the increase of the bending angle is due to 
the increase of overlap between two neutrons
in the $\pi$-orbit and two neutrons in the $\sigma$- orbit.
When there arises the overlap between them, 
the overlap component in the total
wave function is diminished due to Pauli exclusion principle, that is,
so-called Pauli blocking. Therefore, the physical state can be
expressed by the modified wave function which is made by subtracting the
overlap component from the original wave function. Since the energies of 
the $\pi$- and the $\sigma$-
orbits discussed here are relatively low, the modified wave
function involves larger components of higher excitation energy in comparison
with the wave function at $\theta=0$ 
free from the Pauli blocking. As a result, the
Pauli blocking due to the increase of overlap between valence four neutrons
is considered to bring the increase of  energy proportional to the decrease
of squared overlap. This is a possible explanation for rapid increase of the
energy against the bending angle. On this view we can estimate the increase
of the energy as follows.
Firstly, we compare the cases
with Pauli blocking and without Pauli blocking.
If there is no Pauli blocking,
the squared overlap of $^{16}$C$((3/2^-_\pi)^2(1/2^-_\sigma)^2)$
between $\theta = 0^o$ and $\theta = 30^o$ 
is estimated as follows:
\begin{equation}
P(^{16}{\rm C}_{\pi \sigma})_{No Pauli}=
{
P(^{14}{\rm C}_\pi) \times P(^{14}{\rm C}_\sigma) \over P(^{12}{\rm C})
}
= 0.85.
\end{equation}
Here, $P(^{14}{\rm C}_\pi)$, $P(^{14}{\rm C}_\sigma)$, and $P(^{12}{\rm C})$
are squared overlaps 
between $\theta = 0^o$ and $\theta = 30^o$ 
of $^{14}$C$((3/2^-_\pi)^2)$,
 $^{14}$C$((1/2^-_\sigma)^2)$, and $^{12}$C shown in Table I.
In the actual squared overlap 
of $^{16}$C$((3/2^-_\pi)^2(1/2^-_\sigma)^2)$ shown in Table I,
the Pauli blocking effect is automatically included, 
and the value is $P(^{16}{\rm C}_{\pi \sigma})_{With Pauli} = 0.60$.
This difference corresponds to the component excited to higher shells
because of the Pauli blocking effect, and this is 25 $\%$.

\begin{equation}
P(^{16}{\rm C}_{\pi \sigma})_{No Pauli}- 
P(^{16}{\rm C}_{\pi \sigma})_{With Pauli} = 0.25
\end{equation}
Secondly, we estimate the increase of the kinetic energy 
when two valence neutrons in
the $\pi$- or the $\sigma$-orbits are excited to higher shells
by 25 $\%$ due to the Pauli blocking effect.
\begin{equation}
\Delta E = 2\times \hbar \omega \times 0.25 = \sim 10 {\rm MeV}.
\end{equation}
This value is consistent with the increase of the
kinetic energy at $\theta = 30^o$ mentioned above. 
Therefore, we can conclude that this is 
one of the most important mechanisms
which stabilizes the linear-chain state in $^{16}$C.

\section{Summary and discussions}
It is summarized that the linear-chain structure
of $^{16}$C$((3/2^-_\pi)^2(1/2^-_\sigma)^2)$ 
with 3$\alpha$ core 
is the only case to have the simultaneous stabilities
for the breathing-like break up path and for the bending-like path
among $^{12}$C, $^{14}$C, and $^{16}$C.
Other configurations, such as
 $^{14}$C$((3/2^-_\pi)^2)$ and
 $^{16}$C$((3/2^-_\pi)^2(1/2^-_\pi)^2)$ 
are stable against the breathing-like path but not stable
against the bending-like path.
A combination of the $\pi$- and 
the $\sigma$- orbits occupied by four neutrons plays
doubly important roles to make a deep energy pocket for breathing-like path
and to prevent the bending-like free motion of the system. 
The $^{16}$C$((3/2^-_\pi)^2(1/2^-_\sigma)^2)$ configuration 
forms a rotational band with an energy slope
of ${\hbar^2 \over 2I} = 150$ keV,
and since the two-neutron separation-energy of $^{16}$C is 
experimentally found to be 
$5.5$ MeV, the band head energy is expected to corresponds 
to around 25 MeV in excitation.

We have shown in $^{16}$C that the Pauli-blocking effect among
the valence neutrons play an important
role for the stability of the linear-chain configuration.
This effect is expected to be more important as 
the neutron number increases.
We are interested in $^{18}$C
the appearance of similar linear-chain
structure in low-lying region,
where all of the three neutron orbits introduced 
here are occupied ($(3/2^-_\pi)^2(1/2^-_\pi)^2(1/2^-_\sigma)^2$).

Experimentally,
the observation of the $\gamma$-ray
from high-spin states is the most probable way 
to confirm the presence of rotational band
with large moment of inertia.
Although the present analysis 
has been restricted to the $0^+$ state, 
we will investigate also high-spin states,
where it is expected that 
a glue-like role of valence neutrons becomes more important
to prevent the breathing-like break-up of the system.

In the present analysis, parameter $d$ has been fixed 
when the stability against the bending motion has been 
examined, and $\theta$ has been fixed to zero
when the bending motion has been examined.
However,
to assure theoretically
the stability of these linear-chain states,
we intend to investigate the case when two degrees of freedom 
($d$ and $\theta$) simultaneously activate.
This shall be examined
by superposing states on the energy surface of $d$ and $\theta$ 
based on generator coordinate method.

\vspace*{1cm}

The authors thank members of RI beam science laboratory
in RIKEN for discussions and encouragements.
One of the authors (N.I) thanks 
Prof. R. Lovas, Prof. W. von Oertzen,
Prof. H. Horiuchi, and Dr. Y. Kanada-En'yo
for fruitful discussions.

%******************************************************************8
% Fig. 1
%******************************************************************8
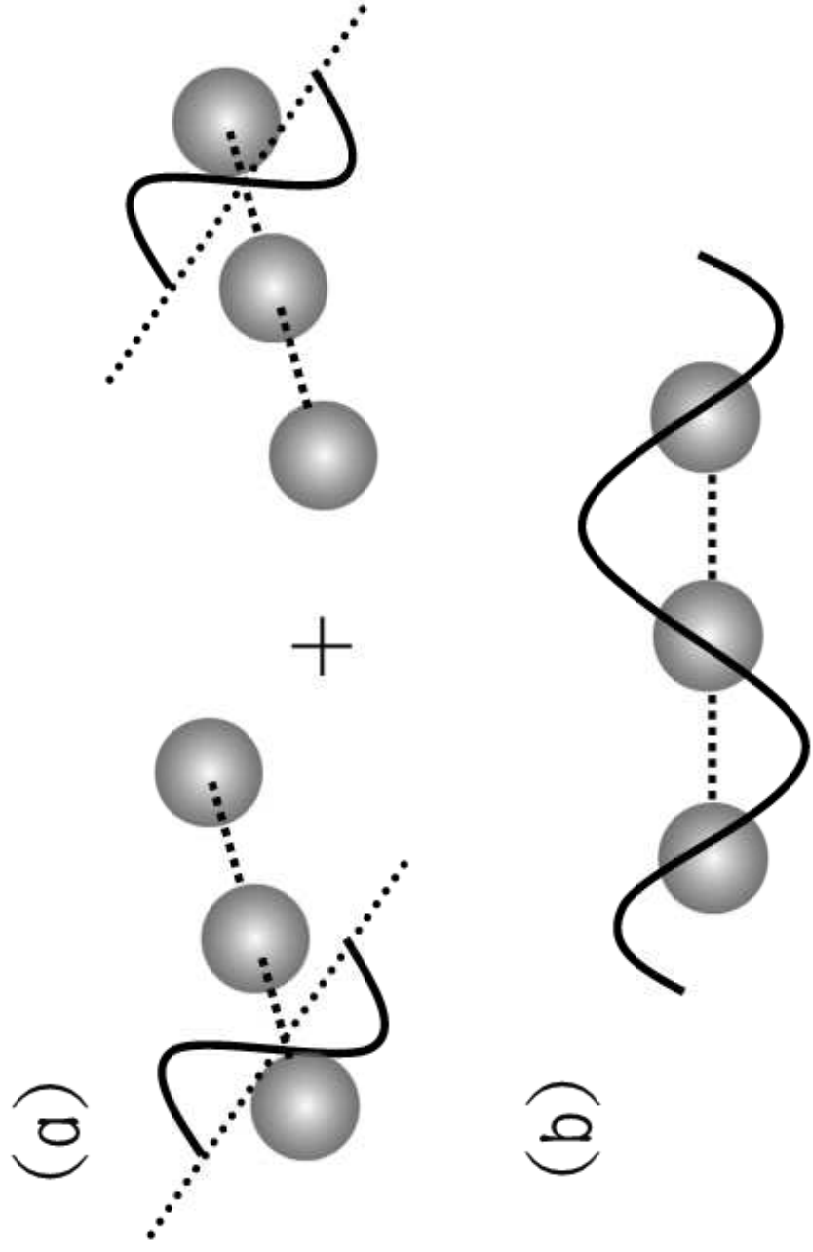
\begin{figure}
%\centerline{\hbox{\psfig{figure=fig1.ps,height=10cm}}}
\caption{The schematic figure for the amplitude 
of the $\pi$-orbit (a) and the $\sigma$-orbit (b) on the $x-z$ 
plane. The circles represent the $\alpha$-clusters.
}
\end{figure}

%******************************************************************8
% Fig. 2
%******************************************************************8
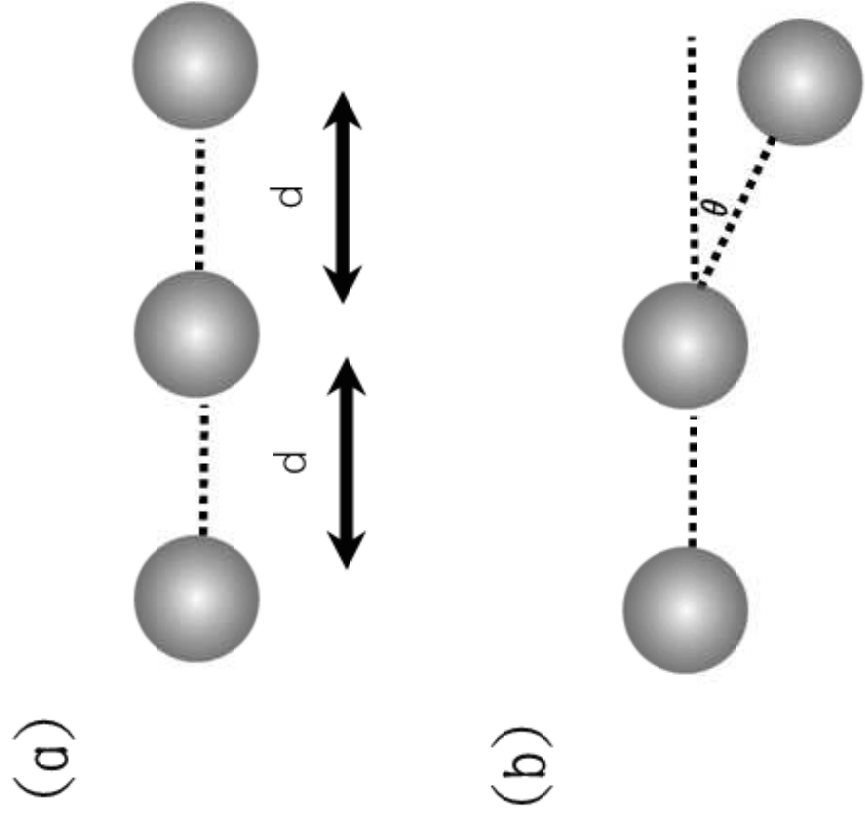
\begin{figure}
%\centerline{\hbox{\psfig{figure=fig2.ps,height=10cm}}}
\caption{
The schematic figure for the breathing (a) and the bending (b) motion
of the linear-chain state.
The stability of the linear-chain state is examined for
the $\alpha$-$\alpha$ distance ($d$ in (a)) and bending angle
($\theta$ in (b)).
}
\end{figure}

%******************************************************************8
% Fig. 3
%******************************************************************8
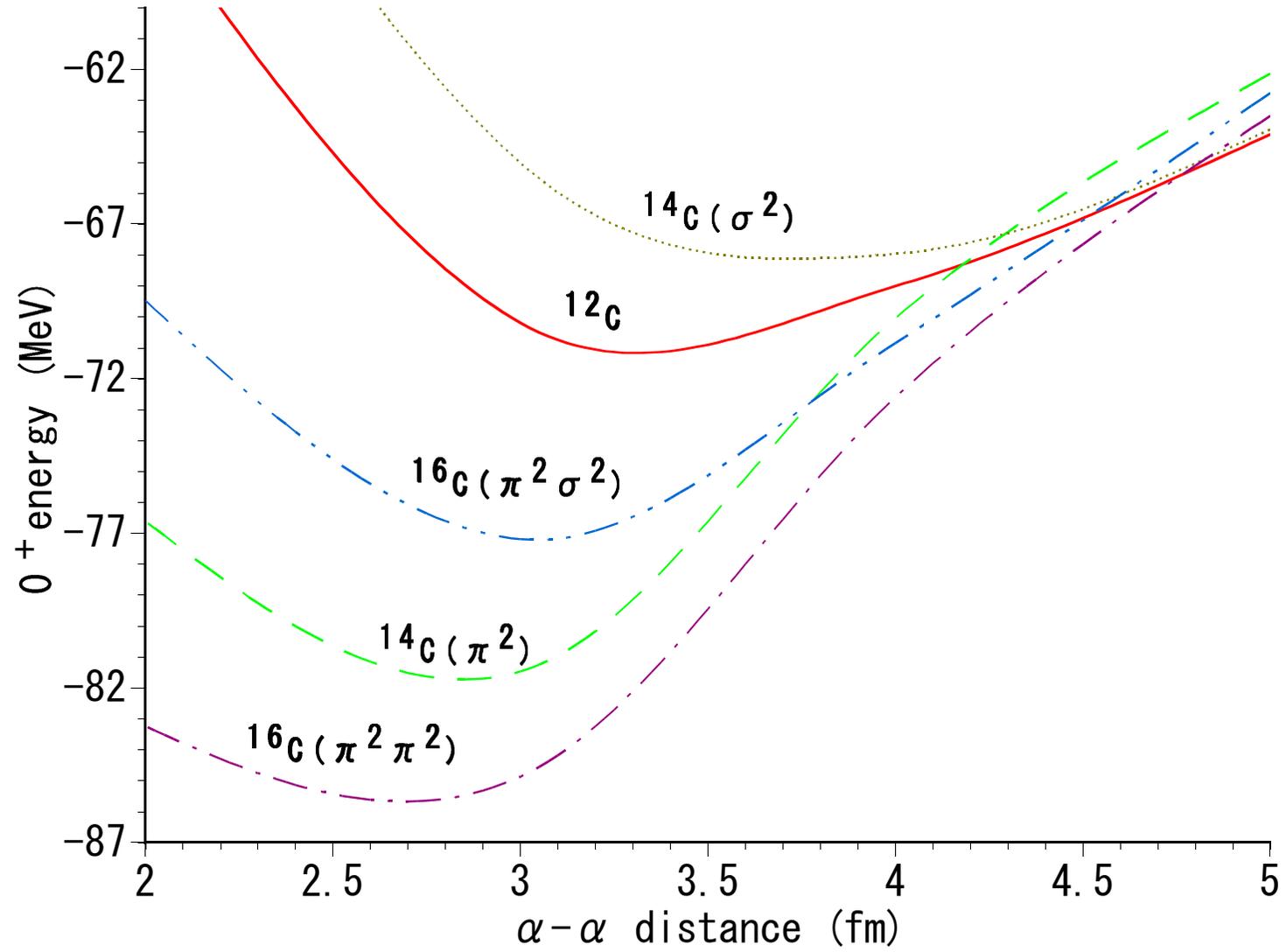
\begin{figure}
%\centerline{\hbox{\psfig{figure=fig3.ps,height=15cm,angle=270}}}
\caption{
The $0^+$ energy curves 
against the $\alpha$-$\alpha$
distance ($d$) 
for $^{12}$C (solid curve),
$^{14}$C(($3/2^-_\pi)^2$) (dashed curve), 
$^{14}$C(($1/2^-_\sigma)^2$) (dotted curve), 
$^{16}$C(($3/2^-_\pi)^2 (1/2^-_\pi)^2)$ (dash dotted curve), 
and 
$^{16}$C(($3/2^-_\pi)^2 (1/2^-_\sigma)^2)$ (dash two-dotted curve).
The coefficients of local Gaussians $\{ g_j\}$ describing 
the valence neutrons are treated as variational parameters
to take into account deviations of the original MO.
}
\end{figure}
%******************************************************************8
% Fig. 4
%******************************************************************8
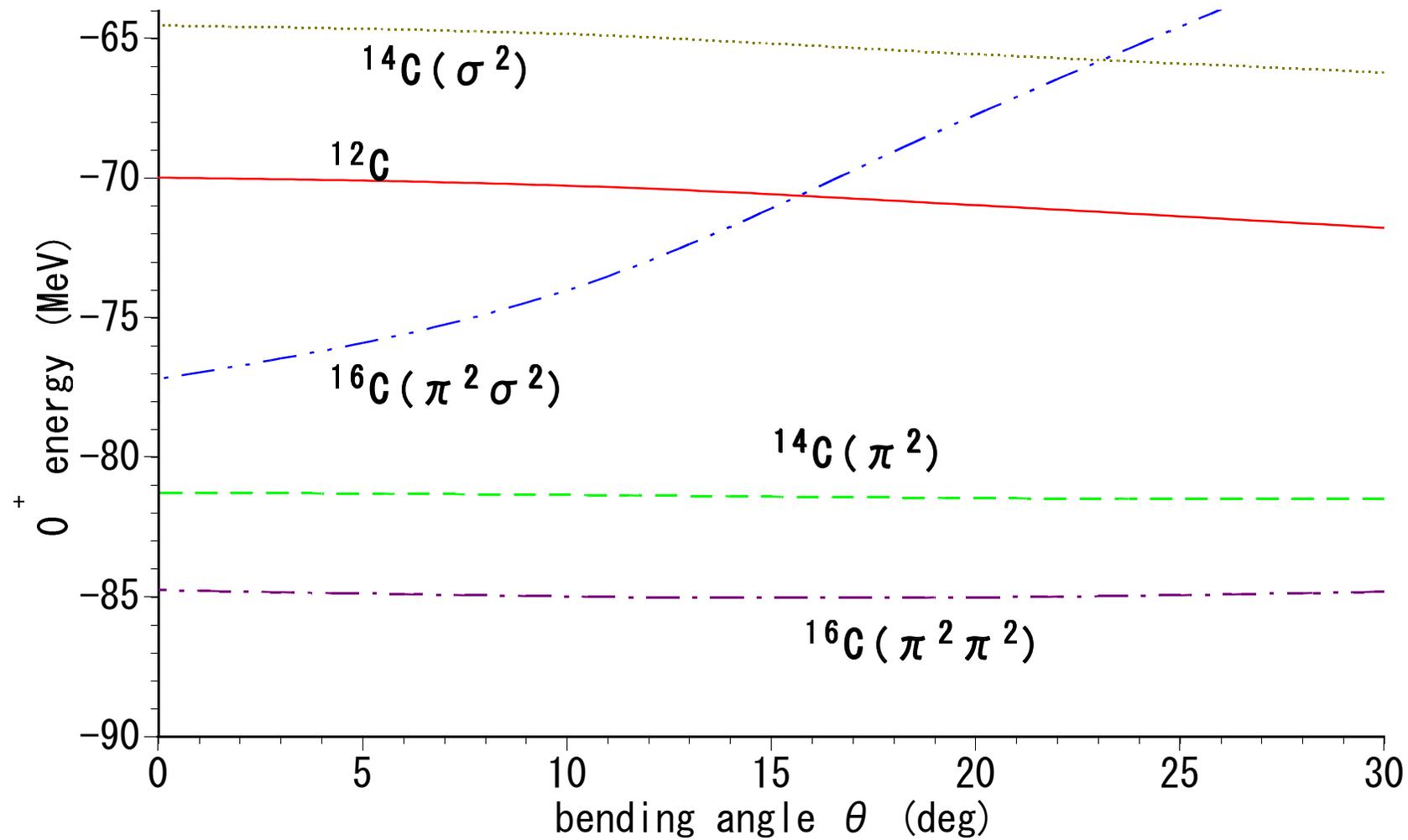
\begin{figure}
%\centerline{\hbox{\psfig{figure=fig4.ps,height=15cm,angle=270}}}
\caption{
The $0^+$ energy curves 
against the bending angle ($\theta$)
for 
$^{12}$C (solid curve), 
$^{14}$C(($3/2^-_\pi)^2$) (dashed curve), 
$^{14}$C(($3/2^-_\sigma)^2$) (dotted curve),
$^{16}$C(($3/2^-_\pi)^2 (1/2^-_\pi)^2)$ (dash dotted curve),
and $^{16}$C(($3/2^-_\pi)^2 (1/2^-_\sigma)^2)$ (dash two-dotted curve). 
The coefficients for the linear combination of Gaussians
describing MO are optimized at $\theta=0^o$ and fixed.
The $\alpha$-$\alpha$ distance ($d$) is fixed to 3 fm.
}
\end{figure}

\begin{table}
\caption{
The squared overlap between $\theta = 0^o$ and $\theta = 30^o$
for
$^{12}$C,
$^{14}$C(($3/2^-_\pi)^2$),
$^{14}$C(($3/2^-_\sigma)^2$),
$^{16}$C(($3/2^-_\pi)^2 (1/2^-_\pi)^2)$,
$^{16}$C(($3/2^-_\pi)^2 (1/2^-_\sigma)^2)$,
}

\begin{center}
\begin{tabular}{|c|c|}
 configuration &  
squared overlap between $\theta = 0^o$ and $\theta = 30^o$ \\ 
\hline
$^{12}$C                                   &   0.906     \\ 
$^{14}$C(($3/2^-_\pi)^2$)                  &   0.905     \\
$^{14}$C(($3/2^-_\sigma)^2$)               &   0.804     \\
$^{16}$C(($3/2^-_\pi)^2 (1/2^-_\pi)^2)$    &   0.865     \\
$^{16}$C(($3/2^-_\pi)^2 (1/2^-_\sigma)^2)$ &   0.602     \\
\end{tabular}
\end{center}
\end{table}


\begin{references}
\bibitem{Soi}
N. Soi\' c $et$ $al$, Eurohys. Lett, {34(1)}, 7 (1996).
\bibitem{Kor}
A. A. Korscheninnikov $et\ al.$
Phys. Lett. B {\bf 343}, 53 (1995).
\bibitem{Freer} M. Freer $et\  al.$,
Phys. Rev. Lett. {\bf 82}, 1383 (1999)
\bibitem{Okabe-S}
H. Furutani, H. Kanada, T. Kaneko, S. Nagata, H. Nishioka, S. Okabe
S. Saito, T. Sakuda, and M. Seya,
Prog. Theor. Phys. Suppl. {\bf 68}, 193 (1980).
\bibitem{Seya}
M. Seya, M. Kohno, and S. Nagata,
Prog. Theor. Phys. {\bf 65}, 204 (1981).
\bibitem{EnyoBe10} Y. Kanada-En'yo, H. Horiuchi and A. Dot{\'e}, 
Phys. Rev. C {\bf 60}, 064304 (1999).
\bibitem{Ogawa} Y. Ogawa, K. Arai, Y. Suzuki, and K. Varga,
Nucl. Phys. A{\bf 673}, 122 (2000).
\bibitem{Ita}
N. Itagaki and S. Okabe, Phys. Rev. C {\bf 61} 044306, (2000).
\bibitem{Ikeda68}
K. Ikeda, N. Takigawa, and H. Horiuchi,
Prog. Theor. Phys. Suppl. Extra Number, 464, (1968).
\bibitem{Morinaga}
H. Morinaga, Phys. Rev. {\bf 101}, 254 (1956); 
Phys. Lett. {\bf 21}, 78 (1966).
\bibitem{Suppl68}
Y. Fujiwara, H. Horiuchi, K. Ikeda, M. Kamimura, K. Kat\=o, Y. Suzuki,
and E. Uegaki,
Prog. Theor. Phys. Suppl. {\bf 68}, 60 (1980).
\bibitem{Cheva}
P. Chevallier, F. Scheibling, 
G. Goldring, I. Plesser, and M. W. Sachs,
Phys. Rev. {\bf 160}, 827 (1967).
\bibitem{Horiuchi72}
H. Horiuchi, K. Ikeda, and Y. Suzuki,
Prog. Theor. Phys. Suppl. {\bf 52}, 89 (1972).
\bibitem{Rae}
W. D. M. Rae, A.C. Merchant, and B. Buck, Phys. Rev. Lett.
{\bf 69}, 3709 (1992).
\bibitem{Oertzen}
W. von Oertzen, Z. Phys. A{\bf 354}, 37 (1996); A{\bf 357}, 355 (1997).
\bibitem{Ita2}
N. Itagaki, S. Okabe, and I. Ikeda, Phys. Rev. C{\bf 62}, 034301 (2000).
\bibitem{Abe}
Y. Abe, J. Hiura, and H. Tanaka, Prog. Theor. Phys. {\bf 49}, 800 (1973).
\bibitem{Enyo95b}
Y. Kanada-En'yo, H. Horiuchi and A. Ono, 
Phys. Rev. C {\bf 52}, 628 (1995).
\bibitem{Ono92}
A. Ono, H. Horiuchi, T. Maruyama and A. Ohnishi,\\
Prog. Theor. Phys. {\bf 87}, 1185 (1992);
Phys. Rev. Lett. {\bf 68}, 2898 (1992).
\bibitem{VolkovInt}
A.B. Volkov, Nucl. Phys. {\bf 74}, 33 (1965).
\bibitem{G3RS} N. Yamaguchi, T. Kasahara, S. Nagata and Y. Akaishi,
Prog. Theor. Phys. {\bf 62}, 1018 (1979).
\end{references}
\end{document}